\documentclass[preprintnumbers, prd, twocolumn, showpacs,floatfix,preprintnumbers,
superscriptaddress, nofootinbib]{revtex4}
\usepackage{graphicx}
\usepackage{epsfig}
\usepackage{bm}
\usepackage{amssymb}
\usepackage{float}
\usepackage{amsmath}
\usepackage{subfigure}
\usepackage{dcolumn}
\usepackage{cancel}
\usepackage[colorlinks]{hyperref}
\usepackage[usenames,dvipsnames]{color}
\hypersetup{
     breaklinks=true,
    pdfstartview={FitH},    
    colorlinks=true,       
    linkcolor=blue,          
    citecolor=red,        
    filecolor=magenta,      
    urlcolor=blue,           
    anchorcolor=green,      
    linktocpage=true
}

\newcommand{\Mpl}{M_{\textrm{Pl}}}
\renewcommand{\(}{\left(}
\renewcommand{\)}{\right)}

\newcommand{\nn}{\nonumber}
\def\al{\alpha}
\def\bet{\beta}
\def\gam{\gamma}

\def\Om{\Omega}
\def\sig{\sigma}
\def\lam{\lambda}
\def\ep{\epsilon}

\def\S{\mathcal{S}}

\def\C{\mathcal{C}}
\def\N{\mathcal{N}}

\def\del{\delta}

\def\doi{http://doi.org}
\def\arxiv{http://arxiv.org/abs}
 \def\t{\tilde}
 \def\e{\mathrm{e}}
\def\r{\mathrm{r}}
\def\g{\mathrm{g}}

\def\m{\mathrm{m}}

\def\s{\mathrm{s}}
\def\d{\mathrm{d}}

\begin{document}

 \title{Quintessential inflation with canonical and noncanonical scalar fields and Planck 2015 results}

 \author{Chao-Qiang Geng}
\email{geng@phys.nthu.edu.tw}
\affiliation{Department of Physics, National Tsing Hua
University,
Hsinchu, Taiwan 300}
 \affiliation{National Center for Theoretical Sciences,
Hsinchu,
Taiwan 300}

 \author{Md. Wali Hossain}
 \email{wali@ctp-jamia.res.in}
\affiliation{Centre for Theoretical Physics, Jamia Millia
Islamia,
New Delhi-110025, India}

\author{R. Myrzakulov}
\email{rmyrzakulov@gmail.com} \affiliation{ Eurasian  Intern
ational
Center for Theoretical Physics, Eurasian National University
, Astana
010008, Kazakhstan}

\author{M.~Sami}
\email{sami@iucaa.ernet.in} \affiliation{Centre for Theoretical
Physics, Jamia Millia Islamia, New Delhi-110025, India}

\author{Emmanuel N. Saridakis}
\email{Emmanuel\_Saridakis@baylor.edu}
 \affiliation{Physics
Division, National Technical University of Athens, 15780
Zografou
Campus,  Athens, Greece} \affiliation{Instituto de F\'{\i}
sica,
Pontificia Universidad de Cat\'olica de Valpara\'{\i}so,
Casilla
4950, Valpara\'{\i}so, Chile}

\begin{abstract}
We investigate two classes of models of quintessential inflation,
based upon canonical as well as noncanonical scalar fields. In
particular, introducing potentials steeper than the standard
exponential, we construct models that can give rise to a successful
inflationary phase, with signatures consistent with Planck 2015
results. Additionally, using  nonminimal coupling of the scalar field
with massive neutrino matter, we obtain the standard thermal history
of the Universe, with late-time cosmic acceleration as the last stage of
evolution. In both cases, inflation and late-time acceleration
are connected by a tracker solution.
\end{abstract}


\maketitle

\section{Introduction}

Theoretical and observational consistency demands that the standard
model of the Universe should be complemented by an early phase of
rapid expansion dubbed inflation
\cite{Starobinsky:1980te,Starobinsky:1982ee,Guth:1980zm,Linde:1983gd,
Linde:1981mu,
Liddle:1999mq,Langlois:2004de,Lyth:1998xn,Guth:2000ka,Lidsey:1995np,
Bassett:2005xm, Mazumdar:2010sa,Wang:2013hva,Mazumdar:2013gya}, as
well as by late time cosmic acceleration
\cite{Peebles:2002gy,Copeland:2006wr,Sahni:1999gb,Padmanabhan:2006ag,Frieman:2008sn,
Cai:2009zp}. The latter is now accepted as a phenomenon of nature
supported by independent
 sets of observations,
whereas inflation still awaits similar confirmation. The relic
gravitational waves generated quantum-mechanically during inflation
would have been a clear and direct signal of inflation
\cite{Grishchuk:1974ny,
Starobinsky:1979ty,Allen:1987bk,Sahni:1990tx,Sahni:2001qp}. In case
the large value of $r$, investigations of B-mode polarization \cite{Ade:2014xna} could
become a powerful tool to falsify the inflationary paradigm.
Unfortunately, the Planck 2015 results \cite{Planck:2015xua,Ade:2015oja}  seem to further shrink the
bound on the tensor-to-scalar ratio of perturbations such that $r=0$
is not ruled out \cite{Ade:2015tva,Planck:2015xua,Ade:2015oja}.

Needless to say that inflation is one of the most beautiful and simple
idea that not only resolves the inconsistencies of the hot big-bang
such as the flatness problem, the horizon problem and others, but
also provides us with a mechanism of generation of primordial
perturbations. As for late time cosmology, the standard model of the
Universe is faced with yet another problem related to the age of the
Universe, which is a late-time phenomenon \cite{Krauss:1995yb,
Turner:1997de,Krauss}. Interestingly, the resolution of the
inconsistency within the framework of standard lore, asks for
late-time cosmic acceleration, which was indeed confirmed directly
by Ia supernovae observations in 1998 \cite{SN11,SN22} and was
indirectly supported by other probes independently
\cite{WMAP2,Ade:2013zuv}. Obviously, accelerated expansion plays an
important role in the history of Universe, both at its early and
late stages.

Often, these two regimes of accelerated expansion are treated
independently. However, it is both tempting and economical to think
that there is a unique cause responsible for both phases of
acceleration {\it \`a la} quintessential inflation
\cite{Peebles:1999fz,Sami:2004xk,
Sami:2004ic,Copeland:2000hn,Huey:2001ae,
Dimopoulos:2000md,Sami:2003my,Dias:2010rg,BasteroGil:2009eb,
Chun:2009yu,Bento:2008yx,Neupane:2007mu,
Rosenfeld:2006hs,Membiela:2006rj,
BuenoSanchez:2006eq,Cardenas:2006py, Zhai:2005ub,
Rosenfeld:2005mt,Giovannini:2003jw,Nunes:2002wz,Dimopoulos:2001ix,
Yahiro:2001uh,Kaganovich:2000fc,Peloso:1999dm,Hossain:2014xha,Hossain:2014coa,
Hossain:2014ova,Hossain:2014zma}, which refers to unification of
both concepts using a single scalar field. Consistency of the
scenario demands that the new degree of freedom, namely the scalar
field, should not interfere with the thermal history of the
Universe, and thereby it should be ``invisible'' for the entire
evolution and reappear only around the present epoch giving rise to
late-time cosmic acceleration. It is, indeed, challenging to build a model
which could successfully comply
 with the said requirements.

First of all, one needs to construct an inflationary phase with a
successful exit. Furthermore, in this scenario one needs an
alternative reheating method (since the  scalar field  must survive
till late times the conventional reheating is not applicable) and
instant preheating \cite{Felder:1998vq,Felder:1999pv,Campos:2004nc} is one of the
efficient
mechanisms that allows
conversion of a part of the scalar field energy into radiation. In
the post inflationary era till the present epoch, the field
potential should be steep, allowing  the radiation domination to
commence, followed by a thermal history as envisaged by hot big
bang. The latter is necessary for sending the field into hiding
after the end of inflation. In particular, the post inflationary dynamics is
characterized by a field that evolves into the kinetic regime for
quite some time, but it then overshoots the background and gets
frozen on its potential due to Hubble damping. As the background
energy density redshifts to the order of the field energy density, the
field resumes its evolution. In case the potential is of a steep
exponential form or steeper, the field tracks the background until
late times \cite{Hossain:2014zma}. In case the potential is
effectively shallow at late times, the field would exit from the
scaling regime to slow roll.

 These features look very viable and pleasing, since it is implied that the late time
evolution is broadly independent of initial conditions. The main reason for demanding
tracker behavior  \cite{Steinhardt:1999nw} after inflation is
related to the hope of alleviation of the fine tuning. However, if
we consider the interaction of the scalar field with matter, the
mass of the scalar is destabilized, bringing back the same level
of fine tuning with the cosmological constant paradigm \cite{Hossain:2014zma}.

In this paper we shall investigate models of quintessential
inflation using canonical (Sec.~\ref{sec:phin}) as well as
noncanonical fields (Sec.~\ref{sec:nc}) with tracking behavior. In
particular, we are interested in constructing models that can
produce a successful inflationary phase (SubSec.~\ref{sec:phin_inf}
for canonical field and SubSec.~\ref{sec:nc_inf} for noncanonical
field), with signatures consistent with the Planck 2015 results, and
then lead to the standard thermal history of the Universe, with
late-time acceleration as the last stage (SubSec.~\ref{sec:phin_late} for canonical field
and SubSec.~\ref{sec:nc_late} for noncanonical field). Finally,  in Sec
\ref{Conclusions} we summarize our results.

\section{Unifying inflation and quintessence using a canonical scalar field}
\label{sec:phin}

In this section we study quintessential inflation using a canonical
scalar field. We consider the action
\begin{eqnarray}
&&\mathcal{S} = \int d^4x
\sqrt{-g}\bigg[\frac{\Mpl^2}{2}R-\frac{1}{2}\partial^\mu\phi\partial_\mu \phi-V(\phi)
\bigg]\nonumber\\
&& \ \ \ \ \ \ +\S_\m+\S_\r  \, ,
 \label{action0}
\end{eqnarray}
with $\Mpl$  the Planck mass,  $\phi$ the scalar field, and $V(\phi)$ its
potential. In the above action we have additionally
 considered the matter and radiation sectors $\S_\m$ and $\S_\r$ respectively. These
sectors can be
neglected at the
 inflationary stage, however they will gradually play an important role, giving
rise to the standard thermal history of the Universe
 and finally to the late-time accelerating phase. As usual, we focus on the
case of a
 flat  Friedmann-Robertson-Walker (FRW) geometry, with metric
\begin{align}
\label{metric0}
 ds^2 = -dt^2 +a(t)^2\delta_{ij} dx^idx^j ~,
\end{align}
where $a(t)$ is the scale factor. Friedman equations are given by
\begin{eqnarray}
 3H^2\Mpl^2 &=& \rho_\m+\rho_\r+\frac{1}{2}\dot\phi^2+V(\phi) \, \nonumber  \\
 \(2\dot H+3H^2\)\Mpl^2&=&-\frac{1}{3}\rho_\r-\frac{1}{2}\dot\phi^2+V(\phi) \, ,
 \label{eq:Friedmann}
\end{eqnarray}
and the equation of motion for the scalar field   has the standard
form
\begin{equation}
 \label{scalareom1}
 \ddot\phi+3H\dot\phi+\frac{\d V}{\d\phi}=0 \, .
 \end{equation}

\subsection{Inflation}
\label{sec:phin_inf}
In what follows, we shall  first analyze the inflationary phase in
this scenario, focusing on the signatures on the observables that
allow for a comparison with the Planck data. As usual, in the
inflationary phase one may neglect $\S_\m$, $\S_\r$ and $\S_\nu$,
and thus the dynamics of inflation, as well as its observational
signatures, are determined solely by the scalar field and its
potential. In particular, given the potential $V(\phi)$, one
introduces the slow-roll parameters
\begin{eqnarray}
\label{eps1}
\epsilon&=&\frac{\Mpl^2}{2}\(\frac{1}{V}\frac{{\rm d}V}{{\rm d}\phi}\)^2\, ,~~\,
\,\\
\eta &=& \frac{\Mpl^2}{V}\frac{{\rm d^2}V}{{\rm d}\phi^2}\,
,
\label{eps2}\\
\xi^2&=&\frac{\Mpl^4}{V^2}\frac{{\rm d}V}{{\rm d}\phi}\frac{
{\rm d}^3V}{{\rm d}\phi^3}\, .
\label{eps3}
\end{eqnarray}
Additionally,  the usual condition for ending  inflation is simply
\begin{eqnarray}
 \epsilon|_{\phi=\phi_{\rm end}}=1,
 \label{condend}
\end{eqnarray}
where the subscript {\it end} represents the value at the end of inflation (we
follow the same convention in the rest of the paper).
The number of e-foldings is calculated through
\begin{eqnarray}
 \N=\int_{t}^{t_{\rm end}}  Hdt'=-\frac{1}{\Mpl^2}\int_{\phi}^{\phi_{\rm end}}
\frac{V(\phi')}{\partial V(\phi')/\partial\phi'}d\phi'.
 \label{Ninflat}
\end{eqnarray}
Hence, observables like the tensor-to-scalar ratio ($r$), the scalar spectral
index ($n_\s$) and its
running ($\al_\s={\rm d}n_\s/{\rm d}\ln k$), can be written
as
\begin{eqnarray}
 r &\approx&16\ep\, ,
 \label{eq:rn}\\
 n_\s &\approx& 1-6\ep+2\eta \, ,
\label{eq:n_s} \\
\al_\s &\approx& 16\ep\eta-24\ep^2-2\xi^2 \, .
\label{eq:xi}
\end{eqnarray}

Keeping in mind the discussion in the introduction, we consider the
potential
 \begin{equation}
 \label{potentialn}
V=V_0\e^{-\lambda \phi^n/\Mpl^n},
\end{equation}
where $V_0$ and $\lam$ are the usual parameters. Note that compared
to standard exponential potential, we have allowed for one more
parameter, namely $n$, which would influence the steepness of the
potential. The case $n=1$ has been extensively studied in the
literature
\cite{Hossain:2014xha,Sahni:2001qp,Copeland:2000hn,Ratra:1989uz,Sami:2002fs,
Lucchin:1984yf} and thus in the following we consider the case
$n\neq1$. Moreover, we consider the cases $n\neq2$ and  $n=2$
 separately, since the corresponding expressions are different in these cases.

\subsubsection{$n\neq2$}

In this case the slow-roll parameters (\ref{eps1})-(\ref{eps3}) have
the following form,
 \begin{eqnarray}
  \label{eq:epn00}
  \epsilon=\frac{1}{2}n^2\lambda^2 \(\frac{\phi}{\Mpl}\)^{2n-2}\,,
 \end{eqnarray}
 \begin{eqnarray}
  \eta=-\Mpl^{2 - 2 n} n \lambda \phi^{n-2}  \left[ \Mpl^n (n-1)
 - n \lambda \phi^n\right]\,,
 \label{eq:epn}
 \end{eqnarray}
 \begin{eqnarray}
 && \xi^2=\Mpl^{4 - 4 n} n^2 \lambda^2 \phi^{2 n-4} \left[\Mpl^{2
n} (n^2-3n+2)\right.\nonumber\\
&&\left. \ \ \ \ \ \  -
   3 \Mpl^n ( n-1) n \lambda \phi^n +
   n^2 \lambda^2 \phi^{2 n}\right]\,,
 \end{eqnarray}
where $\phi$ is the value of the field at the horizon crossing.
Additionally, condition (\ref{condend}) gives
 \begin{eqnarray}
  \phi_{\rm end}= \Mpl   \left(\frac{2}{n^2 \lambda^2}\right)^{\frac{1}{2 n-2}},
  \label{eq:phin_end}
 \end{eqnarray}
and thus from (\ref{Ninflat}) we obtain
 \begin{eqnarray}
 &&\!\!\!\!\!\!\N=\frac{\Mpl^{n-2}}{n\lambda(n-2)}\left(\phi^{2-n}-\phi_{\rm
end}^{2-n}\right)\nonumber\\
&& \, =
 \frac{1}{n\lambda(n-2)}\left[\Mpl^{n-2} \phi^{2-n}-   \left(\frac{2}{n^2
 \lambda^2}\right)^{\frac{2-n}{2 n-2}}\right].
 \label{eq:efold}
 \end{eqnarray}
One can revert this expression in order to get
 \begin{eqnarray}
 \phi= \Mpl\, Q(n,\lambda,\N)
 \label{eq:phiin}
 \end{eqnarray}
 with
 {\small{
 \begin{equation}
\!\!\!\!\!\!\! Q(n,\lambda,\N)=
 \left\{
 n \lambda \!\left[ \!(n-2) \N + n \lambda
     2^{\frac{2 -n}{2 ( n-1)}}
     \left(\!\frac{1}{n^2 \lambda^2}\!\right)^{\frac{n}{2 ( n-1)}}
      \right]
      \right\}^{\frac{1}{2-n}}\!\!\!,
 \end{equation}}}
which allows to eliminate $\phi$ in favor of $\N$ in the
slow-roll parameters. In particular we acquire:
  \begin{eqnarray}
  \label{epsilon22}
  \epsilon= \frac{1}{2}n^2\lambda^2\,Q(n,\lambda,\N)^{2n-2}\,,
 \end{eqnarray}
   \begin{equation}
    \label{eta22}
  \eta= n\lambda \,Q(n,\lambda,\N)^{n-2}\left\{1-n+n\lambda\,
    Q(n,\lambda,\N)^n
            \right\}\,,
 \end{equation}
   \begin{eqnarray}
     \label{xi22}
&&\!\!\!\!\!\!\!\!\!\!\!\!\!\!\xi^2=n^2 \lambda^2
\,Q(n,\lambda,\N)^{2n-4}\nonumber\\
&&\cdot \left\{2 +
   n \left[ n-3 - 3 ( n-1) \lambda\, Q(n,\lambda,\N)^n \right.\right.\nonumber\\
   && \left.\left.\ \ \ \ \ \ \ \ \ \ \ +
      n \lambda^2 \,Q(n,\lambda,\N)^{2 n}
      \right]
                   \right\}\,.
        \end{eqnarray}
 Thus, the tensor-to-scalar ratio ($r$), the scalar spectral
index ($n_\s$) and its
running $\al_\s$, can be calculated straightforwardly as functions of  $n,\lambda,\N$
using  (\ref{eq:rn})-(\ref{eq:xi}).

Let us now use the above expressions to determine for which
combinations of
 $n$, $\lambda$ and  e-folding $ \N$ we obtain values of $ n_\s$ and $r$ in
agreement with the
 Planck 2015 results. In particular, we desire to obtain $n_\s=0.9644\pm0.0049$
 (68 \% confidence level,  Planck TT,TE,EE+lowP) consistent with the Planck 2015 results
\cite{Ade:2015oja}  and $0\leq r \leq 0.149$ (recent joint
analysis of BICEP2/Keck Array and Planck data gives $r_{0.05}< 0.12$
at $95\%$ confidence \cite{Ade:2015tva} and when running of the scalar spectral index is allowed Planck 2015 results give $r<0.149$ \cite{Ade:2015oja} at 95 \% confidence). As a starting point, and
for completeness, we are interested in obtaining $0\leq r \leq 0.149$, and thus describing the limiting cases of both
Collaborations. However, later on we will focus on the low values of
this range, in order to obtain agreement with Planck Collaboration \cite{Planck:2015xua,Ade:2015oja}.

\begin{figure}[h]
\centering
\includegraphics[scale=.65]{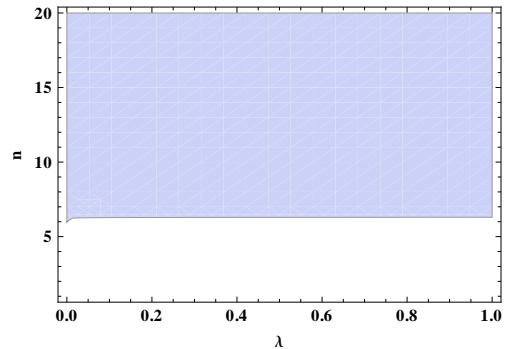}
\caption{{\it{The shaded region marks the allowed  region in the $n-\lambda$
parameter space  that can lead to $n_\s=0.9644\pm 0.0049$ and $0\leq r \leq 0.149$ for $\N=60$.}}}
\label{fig:n_lam}
\end{figure}

\begin{figure}[ht]
\centering
\includegraphics[scale=.65]{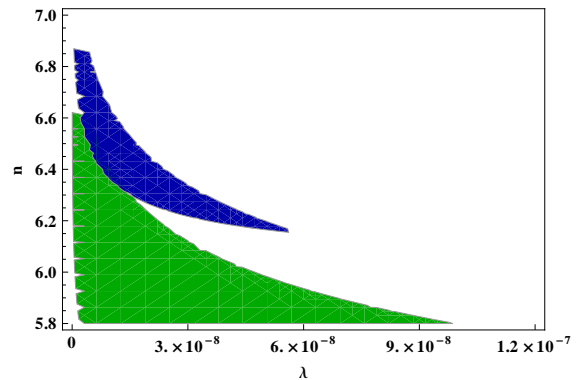}
\caption{{\it{The blue shaded region (upper shaded region) and the green shaded region (lower shaded region) mark the allowed  region in the $n-\lambda$ parameter space  that can lead to $n_\s=0.9644\pm0.0049$  and $0.01\leq r \leq 0.149$ for $\N=60$ and $\N=70$ respectively. }} }
\label{fig:n_lam1}
\end{figure}

 In Fig.~\ref{fig:n_lam} we depict the allowed regions in the $n-\lambda$ parameter space that can give  $n_\s=0.9644\pm0.0049$
\cite{Ade:2015oja} and $0\leq r \leq 0.149$  for $\N=60$. We clearly  see that the parameter $n$ must  be larger than $5$.
It is interesting to notice that if we exclude the zero value, for instance if we
consider $r\geq 0.01$, then the
corresponding region is significantly reduced. In Fig.~\ref{fig:n_lam1} we depict the regions in the $n-\lambda$ parameter space that can give $n_\s=0.9644\pm 0.0049$ and $0.01\leq r \leq 0.149$ for $\N=60$ and $\N=70$, where the aforementioned feature is clear for $\N=60$.  However, this does not seem to be the case according to  both 2013 \cite{Ade:2013uln} and 2015 \cite{Ade:2015oja,Planck:2015xua} Planck data sets. Fig.~\ref{fig:n_lam1} also shows that the parameter space increases if we increase the value of $\N$ from 60 to 70.

\begin{figure}[!]
\centering
\includegraphics[scale=.70]{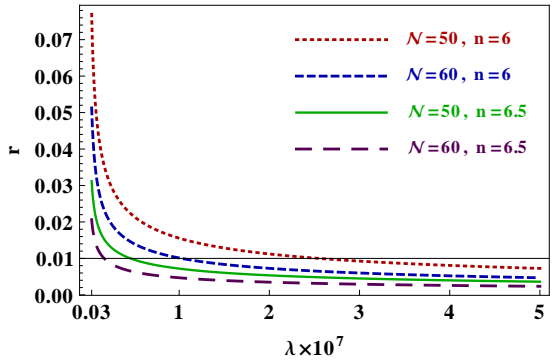}
\caption{\it{{The tensor-to-scalar ratio $r$ versus the parameter $\lambda$, for
different values of the parameter $n$ and e-folding $\N$.}}}
\label{fig:r_lam3}
\end{figure}

\begin{figure}[!]
\centering
\includegraphics[scale=.70]{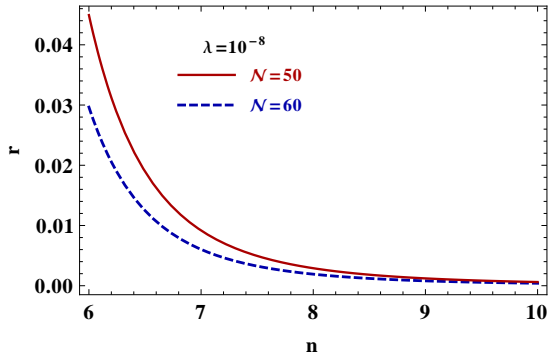}
\caption{
\it{{The tensor-to-scalar ratio $r$ versus the parameter $n$, for fixed parameter
$\lambda$ and different e-folding  value  $\N$.  }}
 }
\label{fig:r_lam4}
\end{figure}

\begin{figure}[!]
\centering
\includegraphics[scale=.70]{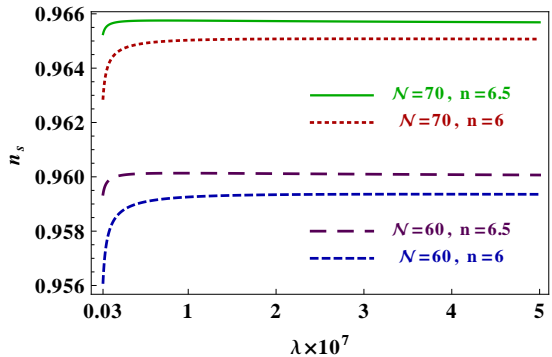}
\caption{\it{{The scalar spectral index  $n_\s$ versus the parameter $\lambda$, for
different values of the parameter $n$ and e-folding $\N$.
}}}
\label{fig:r_lam5}
\end{figure}

\begin{figure}[!]
\centering
\includegraphics[scale=.70]{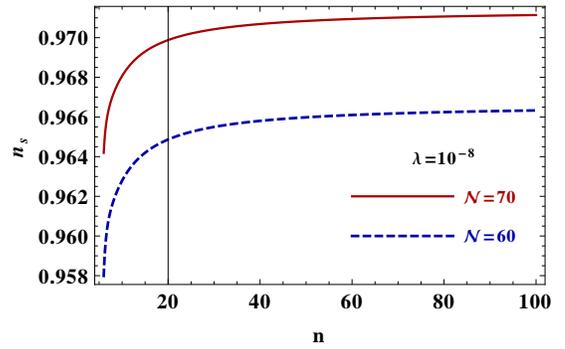}
\caption{\it{{The scalar spectral index  $n_\s$ versus the parameter $n$,  for fixed
parameter
$\lambda$ and different e-folding  value  $\N$.
}}}
\label{fig:r_lam6}
\end{figure}

In order to investigate further the effect of the parameters $n$ and
$\lambda$, for different e-folding number $\N$, on $r$ and $n_\s$,
we include various figures. Firstly, in Fig.~\ref{fig:r_lam3} we
depict $r$ versus   $\lambda$, for different values of   $n$ and
e-folding $\N$, while in Fig.~\ref{fig:r_lam4} we show $r$ versus
$n$, for fixed $\lambda$ and different   $\N$. Similarly, in
Fig.~\ref{fig:r_lam5} we depict   $n_\s$ versus
 $\lambda$ for
different values of   $n$ and  $\N$, while in Fig.~\ref{fig:r_lam6},
we show $r$ versus   $n$, for fixed $\lambda$ and different   $\N$.

\begin{figure}[!]
\centering
\includegraphics[scale=.50]{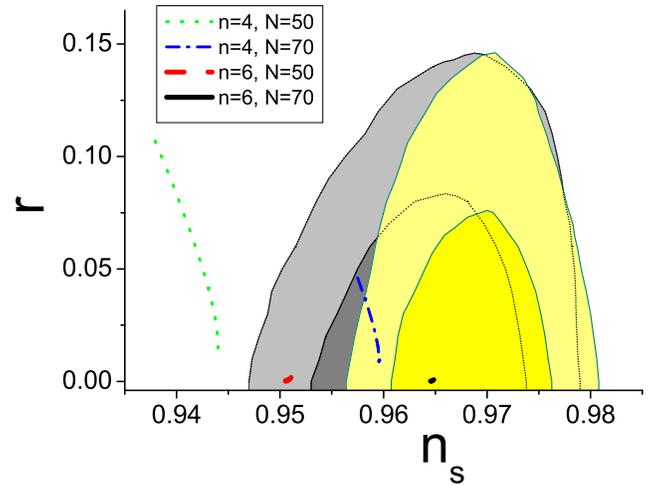}
\caption{{\it{ 1$\sig$ (yellow) and 2$\sig$ (light yellow) contours for Planck 2015
results ($TT+lowP+lensing+BAO+JLA+H_0$)  \cite{Planck:2015xua}, and 1$\sig$ (grey) and
2$\sig$ (light grey) contours for Planck 2013 results  ($Planck+WP+BAO$) \cite{Ade:2013uln} (note that
the  1$\sig$ region of Planck 2013 results  is behind the Planck 2015 results, hence we mark its
boundary by a dotted curve), on $n_s-r$ plane. Additionally, we depict the predictions of
our scenario, for varying $\lambda$ (between  $10^{-6}$ and $10^{-3}$), and $n$ being $4$
or $6$, with the
e-folding value $\mathcal{N}$ being $50$ or $70$.}}}
\label{newPlanckfitcan1}
\end{figure}

\begin{figure}[!]
\centering
\includegraphics[scale=.50]{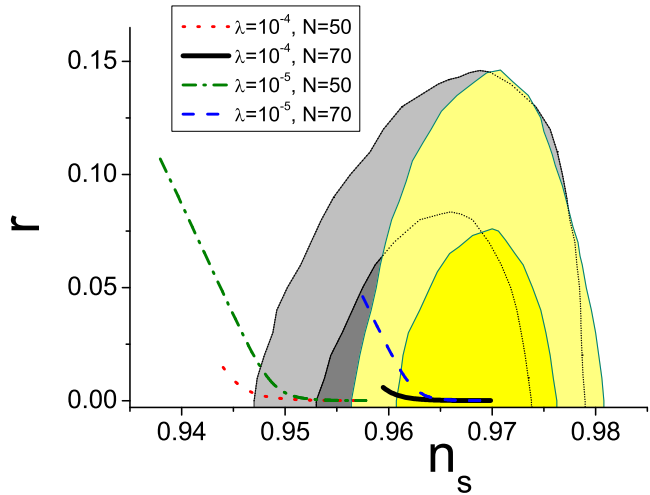}
\caption{{\it{ 1$\sig$ (yellow) and 2$\sig$ (light yellow) contours for Planck 2015 results ($TT+lowP+lensing+BAO+JLA+H_0$)  \cite{Planck:2015xua}, and 1$\sig$ (grey) and
2$\sig$ (light grey) contours for Planck 2013 results  ($Planck+WP+BAO$)  \cite{Ade:2013uln}
(note that
the  1$\sig$ region of Planck 2013 results  is behind the Planck 2015 results, hence we mark its
boundary by a dotted curve), on $n_s-r$ plane. Additionally, we depict the predictions of
our scenario, for varying $n$ (between  $4$ and $20$), and $\lambda$ being
$10^{-4}$ or $10^{-5}$, with the
e-folding value $\mathcal{N}$ being $50$ or $70$.}}}
\label{newPlanckfitcan2}
\end{figure}

It is clear from the above discussion that the scenario at hand,
with the potential (\ref{potentialn}), can give rise to    $n_\s$
and $r$ in agreement with both the Planck 2013 results \cite{Ade:2013uln} and the Planck 2015 results \cite{Planck:2015xua,Ade:2015oja}. In order to present these
features in a more transparent way, in Fig.~\ref{newPlanckfitcan1}
we depict the predictions of our scenario  for varying $\lambda$,
and $n$ being $4$ or $6$, with the e-folding value $\mathcal{N}$
being $50$ or $70$, on top of the 1$\sig$   and 2$\sig$ contours of
the Planck 2013 results \cite{Ade:2013uln} as well as of the  Planck 2015 results
\cite{Planck:2015xua}. As we observe, as $n$ or
$\mathcal{N}$ increase, the predictions move towards the core of the
data.
Furthermore, in Fig.~\ref{newPlanckfitcan2} we present the
corresponding situation, but for varying $n$, and $\lambda$ being
$10^{-4}$ or $10^{-5}$, with   the e-folding value $\mathcal{N}$
being $50$ or $70$. As we observe, as  $\mathcal{N}$ increases the
predictions move towards the core of the data.
Hence, we deduce that  the larger parametric freedom that was introduced by the use of the
additional ``steepness'' parameter $n$, comparing to models with only the parameter
$\lambda$, can lead to the desired $r$-$n_\s$ behavior.

\begin{figure}[!]
\centering
\includegraphics[scale=.47]{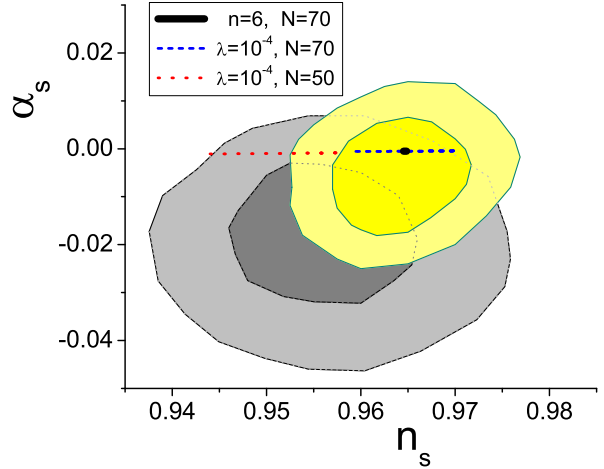}
\caption{{\it{ 1$\sig$ (yellow) and 2$\sig$ (light yellow) contours for Planck 2015
results ($TT,TE,EE+lowP$)  \cite{Planck:2015xua}, and 1$\sig$ (grey) and
2$\sig$ (light grey) contours for   Planck 2013 results  ($\Lambda CDM+running+tensors$) \cite{Ade:2013uln}, on $\alpha_s-n_s$ plane. Additionally, we depict the predictions of
our scenario, for varying $\lambda$ (between  $10^{-6}$ and $10^{-3}$), $n=6$,
$\mathcal{N}=70$, for varying $n$  (between  $4$ and $20$),
$\lambda=10^{-4}$, $\mathcal{N}=70$, and for varying $n$  (between  $4$ and $20$),
$\lambda=10^{-4}$, $\mathcal{N}=50$.}}}
\label{runningspectral}
\end{figure}

For completeness, let us make a comment on the prediction of the scenario at
hand on the running spectral index $\al_\s={\rm d}n_\s/{\rm d}\ln k\approx
16\ep\eta-24\ep^2-2\xi^2$. Using  (\ref{eq:xi}) and
(\ref{epsilon22})-(\ref{xi22}),  we can calculate it for
various values of $\lambda$, $n$ and $\mathcal{N}$, and we present the results on the
$\alpha_s-n_s$ plane in Fig. \ref{runningspectral}. On the same graph we depict the
1$\sig$   and 2$\sig$ contours of the Planck 2013 results  \cite{Ade:2013uln} as well as of the
 Planck 2015 results  \cite{Planck:2015xua,Ade:2015oja}. As we observe, as  $n$ or $\mathcal{N}$
increase the predictions move towards the core of the data, and especially for the
parameter values of Figs. \ref{newPlanckfitcan1} and \ref{newPlanckfitcan2} we obtain a
remarkable agreement with the  Planck 2015 results \cite{Planck:2015xua}.

Let us now calculate the energy scale of inflation using the  COBE normalized
value of density perturbations, which can be represented by the following fitting
function \cite{Bunn:1996py}
\begin{equation}
 \delta_{\rm H}(n_\s,r)=1.91\times 10^{-5}\e^{1.01(1-n_\s)}/\sqrt{1+0.75r} \, .
 \label{eq:delH}
\end{equation}
On the other hand, the scalar perturbation spectrum is
given by
\begin{equation}
 A^2_\s(k)=\frac{V}{\left(150\pi^2\Mpl^4\ep\right)} \, ,
 \label{eq:As}
\end{equation}
and at the horizon crossing ($k=k_*=a_*H_*$) it becomes
\begin{equation}
 A^2_\s(k_*)=7^{n_{s*}-1}\del^2_H \, .
 \label{eq:As_del}
\end{equation}

Using Eqs.~(\ref{eq:delH}), (\ref{eq:As}) and (\ref{eq:As_del}), we
can have the estimation of some model parameters. For instance, for
$r=0.05$ (best fit value of $r$ according to \cite{Ade:2015tva}),
$\N=70$ and $n=6$, equations (\ref{eq:rn}) and (\ref{epsilon22})
give $\lam=1.46\times 10^{-9}$, which leads to $V_0=3.39\times
10^{-9}\Mpl^4$. Additionally, for the same values of $r,\; n,\;
\lam\; {\rm and}\; V_0$, the value of the potential at the
commencement of inflation is $V_{\rm in}=1.4\times 10^{-9} \Mpl^4$,
which provides the scale of inflation as $V_{\rm in}^{1/4}=1.49
\times 10^{16}~\rm GeV$.

Finally, let us discuss on the constraints on reheating temperature from relic
gravitational waves. As shown in Refs.~\cite{Sahni:1990tx,Sahni:2001qp,Hossain:2014xha},
the ratio of the energy densities of relic gravitational waves produced during the
kinetic regime ($\rho_\g$)  and radiation ($\rho_\r$), is
\begin{equation}
 \(\frac{\rho_\phi}{\rho_r}\)_{\rm end}=\frac{3\pi}{64h_{\rm
GW}^2}\(\frac{\rho_\g}{\rho_\r}\)_{\rm
eq} \, ,
\end{equation}
where ``eq'' represents the equality of radiation and scalar field energy densities.
Moreover, the square of relic gravitational wave amplitude writes as
\begin{equation}
 h_{\rm GW}^2=\frac{H_{\rm in}^2}{8\pi \Mpl^2} \, ,
\end{equation}
where $H_{\rm in}$ is the Hubble parameter at the commencement of inflation, which is
$\approx \sqrt{
V_{\rm in}/(3\Mpl)}$.

Nucleosynthesis imposes a constraint on the ratio of the relic gravitational waves and
radiation energy densities, namely $(\rho_\g/\rho_\r)_{\rm eq}\lesssim 0.01$ \cite{Ade:2013zuv}. Hence, this
provides the constraint
on the
amount of radiation energy density at the end of inflation, that is
\begin{eqnarray}
 \rho_{\r,\rm end}\gtrsim
\frac{9V_0^2}{\Mpl^4}~\e^{-\lam\left[Q^n(n,\lam,\N)+\(\frac{\sqrt{2}}{n\lam}\)
^{n/(n-1)}\right]} \, .
 \label{eq:rho_rend}
\end{eqnarray}
Furthermore, the temperature at the end of inflation is $T_{\rm end}=\rho_{\r,\rm
end}^{1/4}$. Therefore, using Eq.~(\ref{eq:rho_rend}) we can also get a constraint on the
temperature at the end of inflation, that is the reheating temperature. If we consider
 $r=0.05$,  $\N=70$ and $n=6$, then we have already seen
that $\lam=1.46\times 10^{-9}$ and $V_0=3.39\times10^{-9}\Mpl^4$. Hence, for these
values of the model parameters we get
\begin{equation}
 T_{\rm end}\gtrsim 2.264\times 10^{14}~ {\rm GeV}\, .
\end{equation}

\subsubsection{$n=2$}

In this paragraph we present the results in the $n=2$ case for
completeness. In this case,  the slow-roll parameters
(\ref{eps1})-(\ref{eps3})  become
\begin{eqnarray}
&&\epsilon=2\lambda^2\frac{\phi^2}{\Mpl^2}\nonumber\\
&&\eta=2\epsilon-2\lambda\nonumber\\
&&\xi^2=4\ep\(\ep-3\lambda\) \, .
\label{eq:slowroll22}
\end{eqnarray}
  Furthermore, the number of e-foldings is
\begin{equation}
\N=\frac{1}{2\lambda}\ln\(\frac{\phi_{\rm end}}{\phi}\),
\end{equation}
with $ \phi_{\rm end}=\frac{\Mpl}{\sqrt{2}\lam}$. Thus,
$\phi$ can be expressed through $\lambda$ and $N$ as
$\phi=\frac{\Mpl}{\sqrt{2}\lam}\e^{-2\lambda \N}
$, and therefore we can write $\ep=\e^{-4\lam \N}$.
Hence, the tensor-to-scalar ratio $r$, the scalar spectral
index $n_\s$ and its
running $\al_\s$, are written as
\begin{eqnarray}
 r &\approx&16\ep=16\e^{-4\lam \N} \, ,
 \label{eq:rb}\\
 n_\s &\approx& 1-6\ep+2\eta=1-2 \e^{-4\lam \N}-4\lam \, ,
\label{eq:n_sb} \\
\lambda_\s &\approx& 16\ep\eta-24\ep^2-2\xi^2 =-8\lam \e^{-4\lam \N} \, .
\label{eq:xib}
\end{eqnarray}

Unfortunately, as one can see expressions (\ref{eq:rb}), (\ref{eq:n_sb}), (\ref{eq:xib})
cannot lead to values in agreement with Planck results for $50\leq\N\leq70$,
independently
of the $\lambda$ value. Thus, we do not investigate this case in more detail.

\subsection{Late Time Dynamics}
\label{sec:phin_late}

\begin{figure}[!]
\centering
\subfigure[]{\includegraphics[scale=.70]{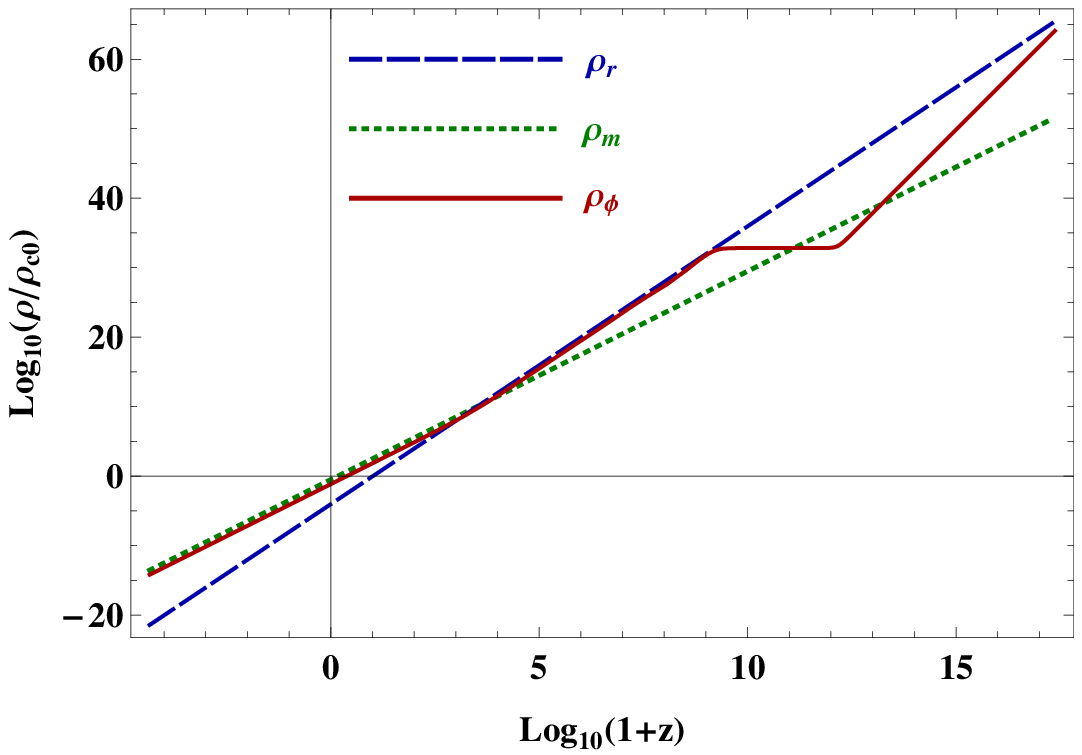}\label{fig:rho_won}}
\subfigure[]{\includegraphics[scale=.70]{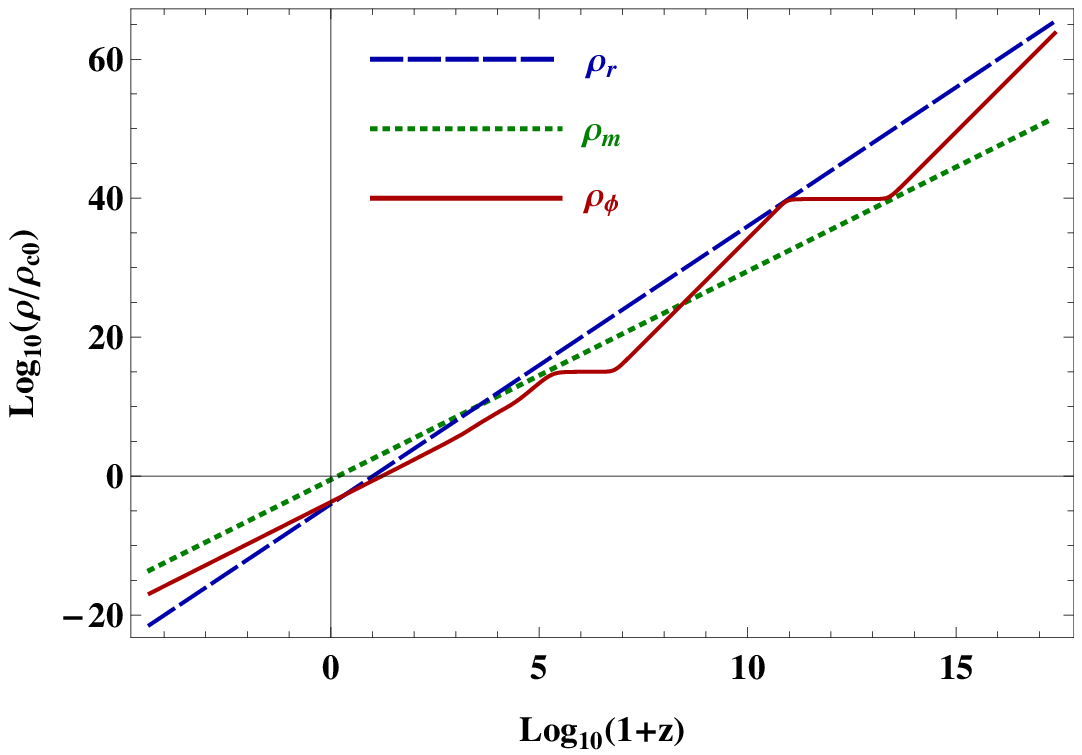}\label{fig:rho_won_n10}}
\caption{{\it{Evolution of the energy densities of matter (dotted green), radiation
(dashed blue) and scalar field (solid red), as a function of the redshift
($z=a_0/a-1$ with $a_0=1$ the present scale factor), in the case of the minimally
coupled scenario (\ref{action0}). The upper graph is for
$n=1$ and  $\lam=4$, while the lower graph is for $n=2.5$ and $\lam=4$.}}}
\end{figure}

In this subsection we investigate the late-time behavior of the above scenario.
For usual steep exponential potential ($n=1$) we know that during the post-inflationary
dynamics the scalar field rolls down the potential, and its energy density  scales as
$\rho_\phi\sim a^{-6}$. Due to the increased Hubble damping, the scalar
field stops evolving, so its energy density eventually becomes comparable to the
background, and it again starts evolving and scales with the background up to late times,
thus leaving no place for late-time acceleration. This class of solutions is known as
scaling solutions \cite{Copeland:1997et}. In Fig.~\ref{fig:rho_won} we present such a
scaling behavior of the scalar field energy density for an exponential potential.

Now let us see what happens when the potential is steeper than the exponential one,
i.e. the case where $n>1$. Similarly to the exponential potential, in this case too the
energy density of the scalar field decreases rapidly, and due to the large Hubble
damping the scalar field  stops  evolving and thus eventually its energy density becomes
comparable with the background one. But unlike the exponential case, now, due
to the very steep nature of the potential, the scalar field cannot follow the
background during the high redshift and thus again $\rho_\phi\sim a^{-6}$. This
results in a rapid decrease in the scalar field energy density and hence again the
scalar field experiences large Hubble damping due to the background, and therefore it
repeats the same behavior as explained earlier (see Fig.~\ref{fig:rho_won_n10}). However,
we mention that this can happen only during large redshifts, where the field value is
not so large.

On the other hand, at late times, when the field evolves to a large
value, the picture is different. In order to understand the behavior
during late times, let us consider the function $\Gamma=V''V/V'^2$.
For an exponential potential $\Gamma=1$, however for the steeper
potential (\ref{potentialn}) we have,
\begin{equation}
 \Gamma=1-\frac{(n-1)}{n\lam}\frac{\Mpl^n}{\phi^n} \, .
 \label{eq:Gam}
\end{equation}
From this expression it is clear that for large $\phi$ and $n>1$ the
function $\Gamma$ approaches 1, i.e., for asymptotically  large
field values, the nature of the potential eventually becomes similar
to the exponential one. Thus, if at late times the field value is
sufficiently large, we obtain a scaling solution, and indeed
Fig.~\ref{fig:rho_won_n10} confirms this.

Unfortunately, as can be also seen in Fig.~\ref{fig:rho_won_n10}, we cannot obtain
late-time acceleration for the potential (\ref{potentialn}). In order to achieve
late-time acceleration  we need a mechanism to exit from the scaling behavior, that is to
obtain the tracker behavior \cite{Steinhardt:1999nw} at late times. For this purpose we
can
consider a nonminimal coupling between the scalar field and massive neutrinos, as in  Refs.~\cite{Hossain:2014xha,Hossain:2014coa,Wetterich:2013jsa} (also see Refs.~\cite{Fardon:2003eh,Bi:2003yr,
Hung:2003jb,Peccei:2004sz,Bi:2004ns,Brookfield:2005td,Brookfield:2005bz,
Amendola:2007yx,Bjaelde:2007ki,Afshordi:2005ym,Wetterich:2007kr,Mota:2008nj,
Pettorino:2010bv,LaVacca:2012ir,Collodel:2012bp}), and we
start with the  action:
\begin{eqnarray}
&&\mathcal{S} = \int d^4x
\sqrt{-g}\bigg[\frac{\Mpl^2}{2}R-\frac{1}{2}\partial^\mu\phi\partial_\mu \phi-V(\phi)
\bigg]\nonumber\\
&&
+\S_\m+\S_\r+\S_\nu\(\C^2g_{\al\bet},\Psi_\nu\) \, ,
 \label{eq:action}
\end{eqnarray}
where
\begin{equation}
 \C^2=A^2 \e^{2\gam\phi/\Mpl} \, .
\end{equation}
In flat  Friedmann-Robertson-Walker (FRW) cosmology the two Friedmann
Eqs.~(\ref{eq:Friedmann})
modify to:
\begin{eqnarray}
 &&3H^2\Mpl^2 = \frac{1}{2}\dot\phi^2+V(\phi)+\rho_\m+\rho_\r+\rho_\nu
 \label{eq:Fried1} \\
 &&\(2\dot H+3H^2\)\Mpl^2 = -\frac{1}{2}\dot\phi^2+V(\phi)-\frac{1}{3}\rho_\r-p_\nu  , \,
~~~~~~
 \label{eq:Fried2}
\end{eqnarray}
while scalar-field equation (\ref{scalareom1}) now becomes
\begin{equation}
 \ddot\phi+3H\dot\phi=-\frac{\d V}{\d \phi}-\frac{\gam}{\
Mpl}(\rho_\nu-3p_\nu) \, .
 \label{eq:eom_phi}
\end{equation}

Before proceeding further we should mention here that massive neutrinos are relativistic ($p_\nu=\rho_\nu/3$) for the most of the expansion history of the Universe and become nonrelativistic ($p_\nu=0$) only after the redshift $z_{\rm NR}\in (2-10)$ for neutrino mass range $m_\nu\in (0.015-2.3)~\rm eV$ \cite{Amendola:2007yx,Wetterich:2007kr}. So concerning the neutrino equation-of-state parameter we shall consider the
following ansatz \cite{Hossain:2014xha}:
\begin{eqnarray}
 w_\nu(z)=\frac{p_\nu}{\rho_\nu}=\frac{1}{6}\left\{1+\tanh
 \left[\frac{\ln(1+z)-z_{\rm eq}}{z_{\rm dur}}\right]\right\} \, ,
 \label{eq:w_nu}
\end{eqnarray}
where $z_{\rm eq}$ and $z_{\rm dur}$ are two parameters which determine the redshift around which the transition of $w_\nu$ from $1/3$ to 0 starts and how fast the transition happens, respectively and the values of these two parameters depend on the redshift $z_{\rm NR}$.

Additionally, the continuity equation for massive neutrinos is given by
\begin{equation}
 \dot\rho_\nu+3H\(\rho_\nu+p_\nu\)=\gam \(\rho_\nu-3p_\nu\)\frac{\dot\phi}{\Mpl} \, .
 \label{eq:cont_nu}
\end{equation}

\begin{figure}[h]
\centering
\subfigure[]{\includegraphics[scale=.6]{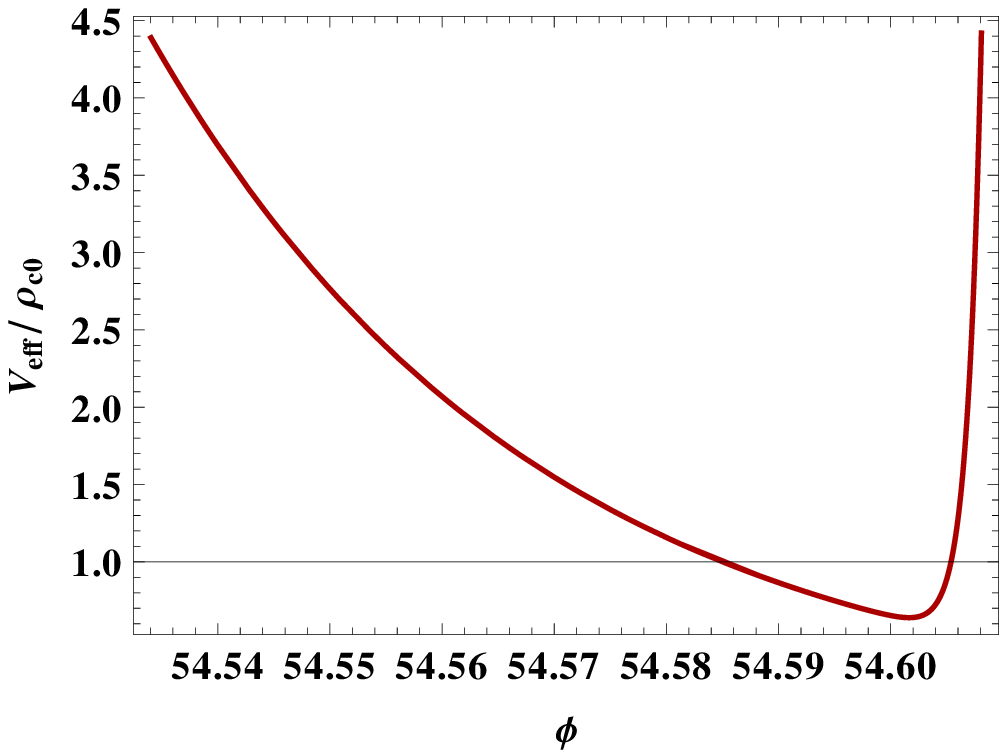}\label{fig:eff_pot}}
\subfigure[]{\includegraphics[scale=.6]{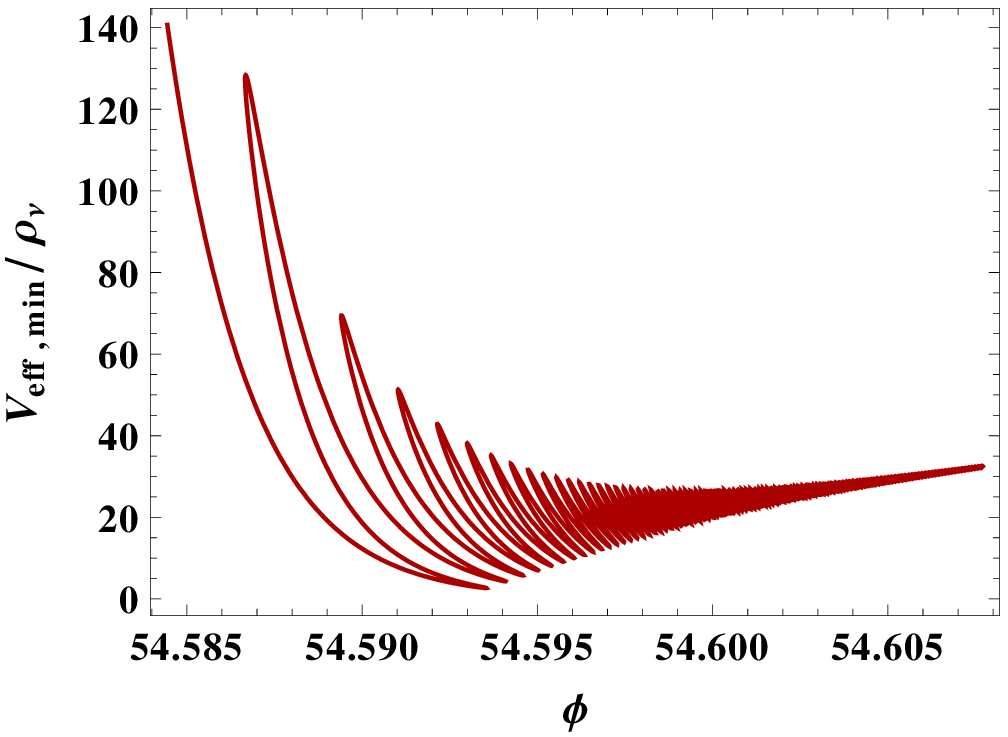}\label{fig:DE_Scale}}
\subfigure[]{\includegraphics[scale=.6]{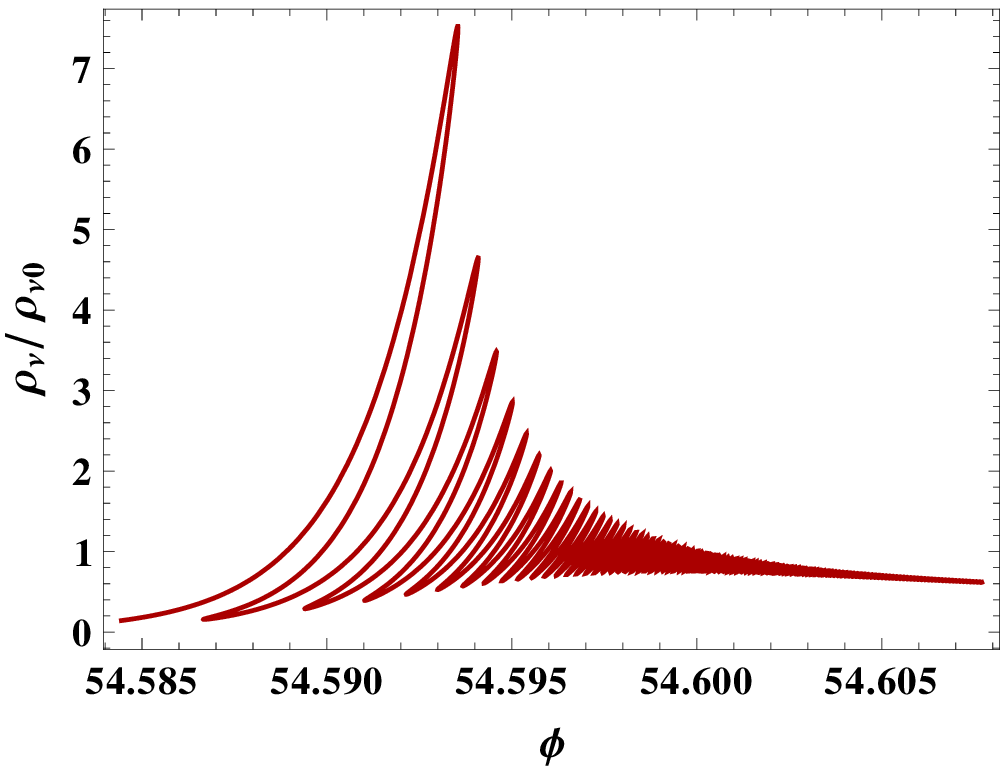}\label{fig:Veff_z}}
\caption{{\it{Top: The minimum of the effective potential (\ref{effpotent})
for $\rho_{\nu0}/\rho_{\rm c 0}=0.0054$ (Planck 2015   results gives
$\Om_{\nu0}h^2<0.0025$ and $H_0=67.74\pm 0.46 ~\rm km~ s^{-1}Mpc^{-1}$\cite{Planck:2015xua}), $\gam=800$ and $\lam=10^{-8}$, with
$\rho_{\rm c 0}$   the present critical density. Middle: The ratio of the minimum of the
effective potential over the massive neutrino energy density versus the field value
around the minimum of the effective potential. Bottom: The neutrino energy
density, normalized with its present value, versus the field value around the minimum
value of the effective potential. For all the plots $\gam=800,\; \lam=10^{-8}\;, n=6\;, z_{\rm eq}=2.55 \; {\rm and}\; z_{\rm dur}=3$. }}}
\end{figure}

The last term of Eqs.~(\ref{eq:eom_phi}) and (\ref{eq:cont_nu}) is effectively zero
when neutrinos behave like radiation, however it becomes non-zero when neutrinos become
nonrelativistic ($p_\nu=0$). So the nonminimal coupling between the massive neutrinos and the scalar field affects the expansion of the Universe only during the late times and as can be deduced from (\ref{eq:eom_phi}), an effective potential forms, which reads as
 \begin{equation}
  V_{\rm eff}=V(\phi)+\rho_{\nu0}\e^{\gam(\phi-\phi_0)/\Mpl}\, ,
  \label{effpotent}
 \end{equation}
where  $\phi_0$ and $\rho_{\nu0}$  are the present values of the
field and of the massive neutrino energy density, and
$\rho_\nu=\rho_{\nu0}\e^{\gam(\phi-\phi_0)/\Mpl}$. This effective
potential clearly has a minimum for $\gam>0$, which forms at late times. Hence,
the scalar field oscillates around this minimum, and eventually it
settles down to the minimum as the oscillations decrease due to the
Hubble friction. For clarity, in Fig.~\ref{fig:eff_pot} we depict
the numerically evolved effective potential, normalized by the
present critical density ($\rho_{\rm c0}$), around its minimum. It should be noted that
the minimum value of the effective potential $V_{\rm eff,min}$ normalized by the present critical density $\rho_{\rm c0}$ is $\approx 1$, which implies that   $\rho_{\rm DE}\approx V_{\rm eff,min}$ since $\rho_{\rm c0}\approx\rho_{\rm DE}$. Moreover,
from Fig.~\ref{fig:DE_Scale} we can see that $V_{\rm eff,min}\sim
\rho_{\rm \nu,min}$, where $\rho_{\rm \nu,min}$ is the energy
density of the massive neutrinos at the minimum of the effective
potential. Furthermore, from Fig.~\ref{fig:Veff_z} we observe that
$\rho_{\rm \nu,min}\approx \rho_{\rm \nu0}$. Hence, in summary we
can deduce that in the model under consideration the dark energy
scale is related to the present energy density of the massive
neutrinos.

\begin{figure}[h]
\centering
\includegraphics[scale=.7]{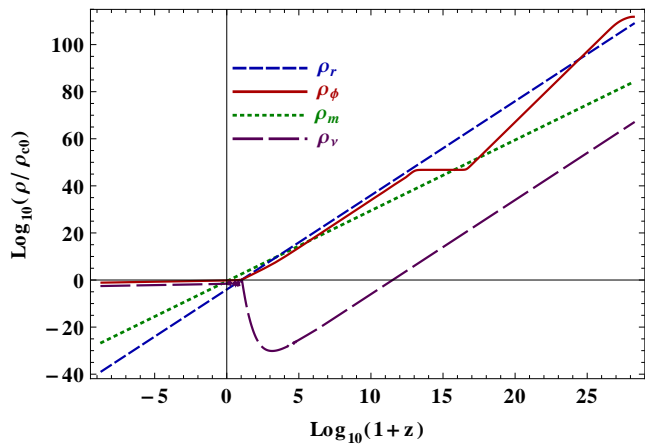}
\caption{{\it{Evolution of the energy densities of matter (dotted green), radiation
(blue short dashed), scalar field (red solid) and massive neutrinos (purple long
dashed), as a
function of the redshift, in the case of
the nonminimally coupled scenario (\ref{eq:action}), for
$\gam=800,\; \lam=10^{-8}\;, n=6\;, z_{\rm eq}=2.55 \; {\rm and}\; z_{\rm dur}=3$. }}}
\label{fig:rho_phin}
\end{figure}

As we described above, the nonminimal coupling between the scalar field and the
neutrinos and the induced effective potential, is adequate to lead to late time
acceleration. Indeed, in Fig. \ref{fig:rho_phin} we depict the evolution of the various
energy densities, and we can clearly see the tracker behavior of the scalar field and the
onset of the dark-energy dominating phase.

\begin{figure}[h]
\centering
\includegraphics[scale=.7]{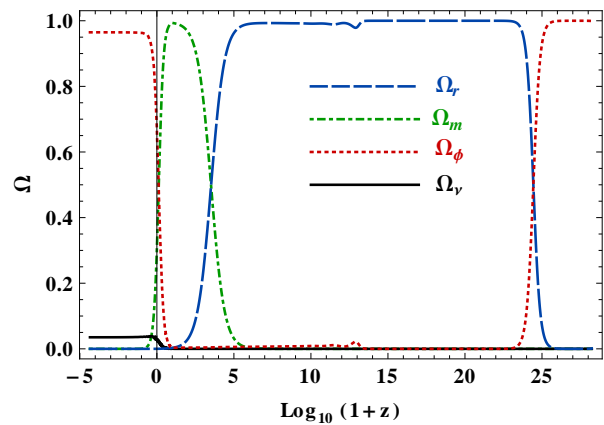}
\caption{{\it{Evolution of the density parameters of radiation (blue dashed), matter
(green dashed-dotted) , scalar field   (red dotted) and neutrinos (black solid),
respectively, in the case of
the nonminimally
coupled scenario (\ref{eq:action}), for   $\gam=800,\; \lam=10^{-8}\;, n=6\;, z_{\rm eq}=2.55 \; {\rm and}\; z_{\rm dur}=3$.
}} }
\label{fig:density_phin}
\end{figure}

\begin{figure}[h]
\centering
\includegraphics[scale=.7]{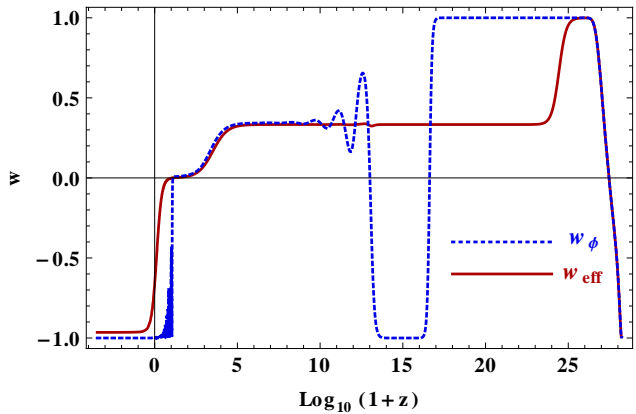}
\caption{{\it{Evolution of the  scalar-field (blue doted)  and  effective (total) (red
solid) equation-of-state parameters, in the case of
the nonminimally
coupled scenario (\ref{eq:action}), for   $\gam=800,\; \lam=10^{-8}\;, n=6\;, z_{\rm eq}=2.55 \; {\rm and}\; z_{\rm dur}=3$.
 }}}
\label{fig:eos_phin}
\end{figure}

In order to present the thermal history of the Universe in a more transparent way, we
introduce the dimensionless density parameters for matter, radiation,
neutrinos and scalar field, respectively given by
\begin{eqnarray}
 \Omega_m &=& \frac{\rho_m}{3H^2\Mpl^2} \, ,
 \label{eq:Omega_m}\\
   \Omega_r &=& \frac{\rho_r}{3H^2\Mpl^2} \, ,
 \label{eq:Omega_r}\\
  \Omega_\nu &=& \frac{\rho_\nu}{3H^2\Mpl^2} \, ,
 \label{eq:Omega_nu}\\
  \Omega_\phi &=& \frac{\rho_\phi}{3H^2\Mpl^2} \, ,
 \label{eq:Omega_sig}
\end{eqnarray}
where $\rho_\phi=(1/2)\dot\phi^2+V$, and in Fig.
\ref{fig:density_phin} we depict their evolution as a function of the redshift.
As we observe, we can reproduce the thermal history of the Universe, starting from a
scalar field kinetic regime, then entering the radiation and matter regimes, and finally
resulting to late-time dark-energy dominated era. Finally, for completeness, in Fig.
\ref{fig:eos_phin} we depict the corresponding behavior of the scalar-field
equation-of-state parameter $w_\phi\equiv p_\phi/\rho_\phi$, as well as of the
effective (total) equation-of-state parameter   $w_{\rm eff}\equiv
p_{\rm tot}/\rho_{\rm tot}=-1-2\dot{H}/3H^2$. From this figure we verify that around the
present era the potential term dominates over the kinetic one in
the scalar-field energy density, which leads $w_\phi$ to be around $-1$.

\section{Unifying inflation and quintessence using a noncanonical scalar field}
     \label{sec:nc}

In this section we shall be interested in constructing
quintessential inflation using a class of models with non-canonical
scalar field \cite{Wetterich:2013jsa,Wetterich:2013aca,Wetterich:2013wza,Wetterich:2014eaa,Wetterich:2014bma,Hossain:2014xha,Hossain:2014zma} (see also Refs. \cite{Nojiri:2005pu,Capozziello:2005tf,Elizalde:2008yf,Ito:2011ae} for unification using noncanonical phantom field). In this case, naturally, we have tracking behaviour in
the post inflationary era \cite{Hossain:2014xha}. Comparing with thawing, the tracking
behavior imposes tough restrictions on the post-inflationary
evolution of the scalar-field dynamics, namely the field should
mimic the background for most of the Universe history, and only at
late times it should exit to slow roll regime. The latter is
realized only for specific potential forms, otherwise the field
exhibits thawing behavior. In summary, in general it is difficult to
acquire the tracking features after inflation, using simple
potentials. However, the picture changes if we use a scalar field
with noncanonical kinetic energy, since in this case it is easy to
control the post-inflationary dynamics in the desired way.

Let us consider the following action \cite{Wetterich:2013jsa,Hossain:2014xha}:
\begin{eqnarray}
\mathcal{S} &=& \int \d^4x
\sqrt{-g}\bigg[\frac{\Mpl^2}{2}R-\frac{k^2(\phi)}{2}\partial^\mu\phi\partial_\mu
\phi-V(\phi) \bigg] \nn \\ &&
+\S_m+\S_r+\S_\nu(\C^2(\phi)g_{\al\beta};\Psi_\nu),
\label{eq:action2}
\end{eqnarray}
with
\begin{eqnarray}
&&k^2(\phi) = \(\frac{\al^2-\t\al^2}{\t\al^2}\)\frac{1}{1+\bet^2\e^{\al\phi/\Mpl}}+1 \,
\label{eq:kphi}\\
&& V(\phi)=\Mpl^4\e^{-\al\phi/\Mpl}
\label{eq:pot_vg_phi}\\
&&  \C(\phi)^2= \zeta \e^{2\tilde{\gamma}\al\phi/\Mpl},
\label{eq:C_vg}
\end{eqnarray}
where $\alpha$, $\tilde{\alpha}$, $\t\gam$ and $\beta$ are the model
parameters. The kinetic function $k^2(\phi)$ has been suitably chosen according to the
tracking post-inflationary requirements.

In order to obtain the realistic intervals of the parameter space,
we start by noting that  $k(\phi)\to 1$ in the large-field limit.
Thus, in this case we obtain a canonical field with exponential
potential, whose slope is given by $\alpha$, which should be large
in order to adhere to nucleosynthesis constraint, namely
$\alpha\gtrsim 20$ \cite{Hossain:2014coa}. Similarly, in the small-field limit we can
introduce a canonical field $\sigma =(\alpha/\tilde{\alpha} )\phi$,
such that the field potential is approximated by $V(\sigma)\sim
e^{-\tilde{\alpha}\sigma/\Mpl} $, and thereby $\tilde{\alpha}$
should be small to comply with inflation \cite{Hossain:2014coa}. Concerning the parameter
$\beta$, it can be fixed by COBE normalization \cite{Hossain:2014coa}. Hence, once the post
inflationary behavior is guaranteed by the specific form of the
kinetic function, inflationary requirements can be obtained through
appropriate choices of $\t\alpha$.

It is clear from the aforesaid that (\ref{eq:action2}) can give rise
to a viable model of quintessential inflation by adding a nonminimal
coupling between the scalar field and the neutrinos described in
subsection \ref{sec:phin_late}. In the following subsection we shall
describe inflation using the noncanonical model (\ref{eq:action2}),
and we will derive constraints on the model parameters in the light
of Planck 2015 results \cite{Planck:2015xua,Ade:2015oja}.
Finally, in a separate subsection we will examine the
late-time, post-inflationary evolution.

\subsection{inflation}
\label{sec:nc_inf}

According to the above discussion, the noncanonical field action (\ref{eq:action2}) can
lead to inflation which commences in the small-field region. In particular,
the slow-roll parameters can be expressed as \cite{Hossain:2014coa}
\begin{eqnarray}
\label{eps1nc}
\epsilon&=&\frac{\Mpl^2}{2k^2(\phi)}\(\frac{1}{V}\frac{{\rm
d}V}{{\rm d}\phi}\)^2
=\frac{\alpha^2}{2k^2(\phi)}\simeq \frac{\tilde{\alpha}^2}{
2}\left(1+X\right) ,~~\,\,\\
\eta &=&  2\epsilon-\frac{\Mpl}{\alpha}\frac{{\rm d}\ep(\phi)}{{\rm
d}\phi}\simeq \epsilon+\frac{\tilde{\alpha}^2}{2} ,
\label{eps2nc}\\
\xi^2&=& 2\ep\eta-\frac{\al\Mpl}{k^2}\frac{{\rm d}\eta}{{\rm
d}\phi}\simeq 2\t\al^2\ep \, ,
\label{eps3nc}
\end{eqnarray}
with $X\equiv \beta^2e^{\alpha \phi/\Mpl}$,  and where we have
used the approximations $\alpha\gg 1$ and $\tilde{\alpha}\ll 1$, which
should hold in the scenario under consideration. It is clear from (\ref{eps2nc})
that inflation ends in the region with $X\gg 1$, which quantifies the
large-field approximation. In this approximation the
number of e-foldings  becomes \cite{Hossain:2014coa}
\begin{equation}
 \N\approx \frac{1}{\tilde\alpha^2}\ln\left(1+X^{-1}\right),
\end{equation}
and thus it can be related to $\epsilon$ through \cite{Hossain:2014coa}
\begin{equation}
 \ep(\N)=\frac{\t\al^2}{2} \frac{1}{1-\e^{-\t\al^2\N}}.
\end{equation}
Let us note that the transition between small and large field regimes takes place when
$\t\al^2\approx 1/\N$. Since inflation always ends in the region of large field, its
commencement depends upon the range of inflation, which is in turn uniquely specified by
the tensor-to-scalar ratio. In particular, a large value of $r$,  or  weak slow roll,
would imply large field excursion. In that case inflation should commence around
the boundary of transition, otherwise the commencement would be
shifted to the large-field region.

\begin{figure}[!]
\centering
\includegraphics[scale=.50]{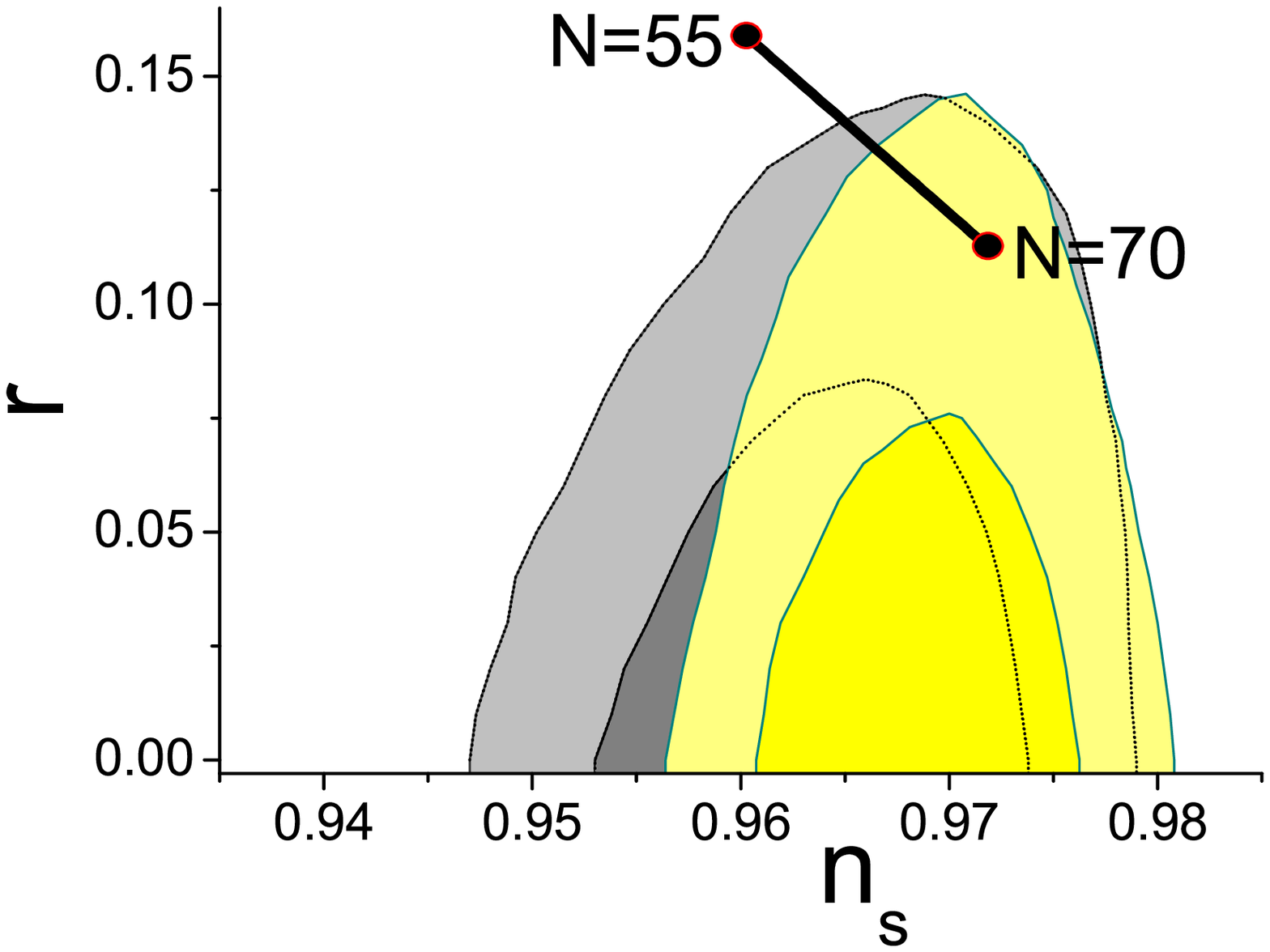}
\caption{{\it{
1$\sig$ (yellow) and 2$\sig$ (light yellow) contours for Planck 2015
results ($TT+lowP+lensing+BAO+JLA+H_0$)  \cite{Planck:2015xua,Ade:2015oja}, and 1$\sig$ (grey) and
2$\sig$ (light grey) contours for Planck 2013 results  ($Planck+WP+BAO$)  \cite{Ade:2013zuv}
(note that
the  1$\sig$ region of Planck 2013 results  is behind the Planck 2015 results, hence we mark its
boundary by a dotted curve), on $n_s-r$ plane.  Additionally, we depict the predictions of
our scenario given by (\ref{eq:r}),(\ref{eq:n_s}), for $\tilde{\alpha}\rightarrow0$ and
e-folding  $\mathcal{N}$ varying between 55 and 70.}}}
\label{newPlanckfitnoncan}
\end{figure}

The general expressions for $r$, $n_\s$ and the running of spectral index  $\al_\s$, valid
from small to large
field regimes are given by
\begin{eqnarray}
&&\!\!\!\!\!\!\!\!\!\!\!\!\!\!
r(\mathcal{N},\t\al)=16\epsilon(\mathcal{N})\approx\frac{
8\t\al^2}{1-\e^{-\tilde\alpha^2\mathcal{N}}} \, ,
 \label{eq:r}\\
&&\!\!\!\!\!\!\!\!\!\!\!\!\!\! n_\s(\N,\t\al)= 1-6\ep+2\eta
\approx 1-\tilde{\alpha}^2\coth\(\frac{\t\al^2\N}{2}\)  \, ,
\label{eq:n_s} \\
&&\!\!\!\!\!\!\!\!\!\!\!\!\!\!
\al_\s\equiv\frac{{\rm d}n_\s}
{{\rm d}
\ln k} = 16\ep\eta-24\ep^2-
2\xi^2\approx
-\frac{\t\al^4}{2\sinh^2\(\frac{\t\al^2\N}{2}\)} .
\label{eq:xic}
\end{eqnarray}
Hence, let us use these expressions in order to compare the predictions of the scenario
at hand with the 2013 and Planck 2015 results. In Fig.
\ref{newPlanckfitnoncan} we depict the predictions of our model    for
$\tilde{\alpha}\rightarrow0$ and
e-folding  $\mathcal{N}$ varying between
  55 and 70, on top of the 1$\sig$   and
2$\sig$ contours of the Planck 2013 results  \cite{Ade:2013zuv} as well as of the Planck 2015 results  \cite{Planck:2015xua,Ade:2015oja}. As we observe, the
point for $\mathcal{N}=55$ lies outside the
2$\sig$ contour of both Planck data sets, but higher values of $\mathcal{N}$
lie within the 2$\sig$ contour of both Planck data. Unfortunately, all predictions
still lie  outside the  1$\sig$ contour of both data sets, and this becomes worse
for larger values of $\tilde{\alpha}$.

As our potential (\ref{potentialn}) is of unusual form, a comment about the viability of model under quantum correction is in order.
  Indeed, since, $V\sim
\e^{\phi^n}, n>5$, the model would involve operator of dimensions
higher than four if we imagine the series expansion of the
potential. It then obviously  raises the question whether
 the model would make sense if quantum corrections are invoked. In
general the effective Lagrangian, after we fix the ignorance,
contains both renormalizable as well as non-renormalizable parts. As
for renormalizable part, it includes one loop corrections to
classical Lagrangian. In our case, the latter is absent. The
non-renormalizable part includes correction that are suppressed
by inverse powers of the cut off. As long as we work quite below the
cut off, we can safely use the classical framework. The effective
Lagrangian in our case has the form,
\begin{equation}
\mathcal{L}=\mathcal{L}_{cl}+\sum_{i=1}^\infty\left(\frac{c_i
\phi^{ni}}{\Lambda^{-4+ni}}+\frac {d_i (\partial
\phi)^2\phi^{ni-4}}{\Lambda^{-4+ni}}\right)
\end{equation}
where we imagined series expansion of the potential and $c_i, d_i$
are constants. In this case, marginal and relevant operators are
absent. Thus, in case we work well below the cut off, we can safely
ignore the correction and keep using the classical framework.

\subsection{Late Time Dynamics}
\label{sec:nc_late}

In the noncanonical scalar-field scenario at hand, after the end of
inflation the Universe enters into a kinetic-energy-dominated
regime, known as ``kinetic regime'', and then it subsequently enters
into the radiation, matter and dark-energy  eras
\cite{Hossain:2014xha}. The nonminimal coupling between the scalar
field and massive neutrinos plays the main roll for the onset of
late-time cosmic acceleration, as we analyzed in subsection
\ref{sec:phin_late}. In particular, when the massive neutrinos
become nonrelativistic at late times, they contribute to the
formation of the effective potential with a minimum. As a result,
the scalar field settles down to the minimum of the effective potential after damping of
oscillations which ultimately gives rise to the late time acceleration. For a
detailed dynamical analysis one can see Ref.~\cite{Hossain:2014xha}.
The late-time attractor solution corresponds to an effective (total)
equation-of-state parameter given by \cite{Hossain:2014xha}
\begin{equation}
 w_{\rm eff}=-\frac{\t\gam}{1+\t\gam} \, ,
 \label{eq:eff_eos_nc}
\end{equation}
and to a scalar-field equation of state
\begin{equation}
 w_\sig=-\frac{\al^2\t\gam(1+\t\gam)}{3+\al^2\t\gam(1+\t\gam)} \, ,
 \label{eq:eos_sig_nc}
\end{equation}
where $\sig$ is the canonical scalar field which can be represented in terms of the
noncanonical
scalar field $\phi$ using the transformation \cite{Hossain:2014xha}
\begin{eqnarray}
\label{sigmadef}
 \sigma &=& \Bbbk(\phi) \, ,\\
 k^2(\phi) &=& \(\frac{\partial\Bbbk}{\partial \phi}\)^2 \, .
 \label{eq:dsig}
\end{eqnarray}
From (\ref{eq:eos_sig_nc}) we can see that if $\t\gam=0$,  i.e. without a coupling
between the scalar field and massive neutrinos, $w_\sig=0$, which implies that the
scalar field will exhibit scaling behavior even during late times, and will continue to
follow the background even in the future. Hence, we do verify what we discussed earlier,
namely  that in the absence of the nonminimal coupling we cannot acquire late-time
acceleration. On the other hand, from (\ref{eq:eff_eos_nc}) we deduce that in order to
obtain a de Sitter or nearly de Sitter solution ($w_{\rm eff}\approx -1$) we require
$\t\gam\gg 1$. Thus, a large nonminimal coupling is needed in order to acquire late-time
acceleration.

Finally, note that the value of the effective potential at the minimum is directly
proportional to the present neutrino energy density
\cite{Hossain:2014xha,Hossain:2014zma}. Therefore, the dark energy scale
is related to the neutrino energy scale, similarly to the analysis of
Section \ref{sec:phin}. Hence, the nonminimal coupling between the scalar field and
massive neutrinos not only provides the late-time acceleration, but it additionally
fixes the energy scale of dark energy.

\section{Conclusions}
\label{Conclusions}

In this work we investigated two distinct classes of quintessential
inflation, namely models based on a canonical scalar-field and
models based on a noncanonical scalar field, where both scenarios
exhibit tracking  behavior. In both cases we considered a nonminimal
coupling between the scalar field and the neutrinos, which is
required in order to trigger the late-time cosmic acceleration.

In the canonical case we considered a potential of the form $V\sim
e^{\lambda \phi^n/\Mpl^2}$, with $n>1$, which has the property of
slow roll near the origin but it becomes steep away from it. Hence,
at early times, i.e. while the field is around the origin, this
scenario can give rise to inflation. Indeed, as we demonstrated in
Figs.~\ref{fig:n_lam} and \ref{fig:n_lam1}, for a range of the model
parameters we can obtain  a required phase of inflation with the
spectral index $n_\s$ and the tensor-to-scalar ratio  $r$ in a very
good agreement with the   Planck 2015   results \cite{Planck:2015xua,Ade:2015oja}
and the joint analysis of BICEP2/Keck Array and Planck data
\cite{Ade:2015tva}. Indeed, in Fig.~\ref{newPlanckfitcan1} and
Fig.~\ref{newPlanckfitcan2} we showed that the predictions of the
scenario fall well inside the 1$\sigma$ likelihood contours of both Planck 2013 results \cite{Ade:2013uln} and Planck 2015 results \cite{Ade:2015oja,Planck:2015xua}. Additionally, for the representative case
of the parameter choice $n=6$, $\lam=1.5\times 10^{-9}$ and
e-foldings $\N=70$, the obtained $r=0.05$ provides the estimate
$V_0=3.39\times 10^{-9}\Mpl^4$ or equivalently the scale of
inflation, $V_{\rm in}^{1/4}=1.49 \times 10^{16}~\rm GeV$. Finally,
using nucleosynthesis constraints we obtained the lower bound on the
temperature at the end of inflation, namely $T_{\rm end}\simeq
2.264\times 10^{14}~\rm GeV$.

After the end of inflation, the steep  potential derives the scalar
field into the kinetic regime. Consequently, the field overshoots
the background and freezes due to Hubble damping. The evolution of
the field resumes soon after the background energy density becomes
comparable to field energy density. In the usual case of exponential
potential ($n=1$) the field follows scaling behavior. In the case of
$n>1$, the potential is steeper than the standard exponential and
thus the field is driven away from the scaling track, which
increases the Hubble damping leading to the freezing of the field
once again. As the field comes out of the freezing regime, its
energy density redshifts faster than the background, and this
feature brings back the Hubble damping and so on. The said behavior
keeps repeating till the field acquires large values. In that case
$\Gamma\to 1$ and the field enters the regime which is an attractor,
see Fig.~\ref{fig:rho_won_n10}. The latter allows us to obtain the
subsequent radiation and matter regimes. Thus, we conclude that the
scaling solution is also an attractor in case the potential is
steeper than a standard steep exponential. To the best of our
knowledge, this feature was not noted earlier in the literature. In
case of generic values of $n$ and $\lambda$, the field rolls in the
domain of large $\phi$ in the post inflationary era in which case
the system enters into the tracking regime after the Hubble freezing
ends, see Fig.~\ref{fig:rho_phin}.

 Finally, the nonminimal coupling between the scalar
field and the neutrinos induces an effective potential, which leads
the scalar field to drive the late-time acceleration. The larger the
nonminimal coupling is, the deeper the minimum of the effective
potential is, in which the field is settled after damped
oscillations, exhibiting an equation-of-state parameter around $-1$
in the present epoch.

As for the noncanonical scalar field, although both inflation and
the subsequent thermal history of the Universe, including late-time
acceleration, can be obtained, the specific values of the spectral
index and of the tensor-to-scalar ratio are not in complete agreement with
the  Planck 2015   results \cite{Planck:2015xua,Ade:2015oja}.

In this work we showed that using potentials steeper than the
exponential we can solve the problem of models of quintessential
inflation  which give rise to numerical values of $r$ larger than
the Planck bounds \cite{Planck:2015xua,Ade:2015oja}, and we can obtain a
remarkable agreement with the  Planck 2015   results
\cite{Planck:2015xua,Ade:2015oja}. We have shown that it is possible to
reproduce the correct post-inflationary evolution during  radiation
and matter eras. The nonminimal coupling between the scalar field
and the neutrinos is shown to drive the late-time cosmic
acceleration. To summarize, we presented a successful model of
quintessential inflation that can describe the entire  history of
Universe evolution in a unified framework.

\begin{acknowledgments}
M.W.H. acknowledges CSIR, Govt. of India for financial support through SRF scheme (File
No:09/466(0128)/2010-EMR-I).  The research of ENS is implemented within the framework
of the Action ``Supporting Postdoctoral Researchers'' of the Operational Program
``Education and Lifelong
Learning'' (Actions Beneficiary: General Secretariat for Research and Technology), and is
co-financed by the European Social Fund (ESF) and the Greek State.
\end{acknowledgments}

\end{document}